\documentclass[11pt, a4paper]{article}
\usepackage{latexsym,amsmath,amssymb,amsbsy,amsthm}

\newtheorem{theorem}{Theorem}[section]
\newtheorem{definition}{Definition}[section]
\newcommand{\refn}[1]{(\ref{#1})}
\newcommand{\tsp}{\mathfrak E}
\newcommand{\hil}{\mathcal H}
\newcommand{\ph}{\varphi}
\newcommand{\phx}{\varphi(t, x)}
\newcommand{\phy}{\varphi(t, y)}
\newcommand{\pix}{\pi(t, x)}
\newcommand{\piy}{\pi(t, y)}
\newcommand{\der}{\partial}

\newcommand{\herm}{h_{n}(x)}
\newcommand{\re}{\mathbb R}
\newcommand{\al}{\alpha}
\newcommand{\be}{\beta}
\newcommand{\kr}{\mathcal K}
\newcommand{\psikr}{\Psi^{(k)}}
\newcommand{\tps}{\tilde\Psi}
\newcommand{\tp}{\tilde P}
\newcommand{\tq}{\tilde Q}
\newcommand{\ident}{\,\textup I}
\newcommand{\ins}{\mathfrak S}

\begin{document}

{\Large\bf HAAG'S THEOREM IN THE THEORIES WITH NON-PHYSICAL PARTICLES}

\medskip

\begin{quotation}\noindent
K.~V.~ Antipin,$^{1}$ M.~N.~Mnatsakanova,$^{2}$ and Yu.~S.~Vernov$^{3}$\vspace{-0.3\baselineskip}
\bigskip

\noindent$^1$\small{\em Department of Physics, Moscow State University, Moscow 119991, Russia.} \\
$^2$\small{\em Skobeltsyn Institute of Nuclear Physics, Moscow State University, Moscow 119992, Russia.} \\
$^3$\small{\em Institute for Nuclear Research, Russian Academy of Sciences, Moscow 117312, Russia.}

\bigskip\bigskip\bigskip\medskip

\noindent Haag's theorem is extended to the case of regular representations of the canonical commutation relations in a non-degenerate indefinite inner product space.
\bigskip
\medskip

\medskip


\end{quotation}
\vspace{-0.5\baselineskip}

\section{Introduction}
\mbox{}\vspace{-\baselineskip}

It is well known that the so-called non-physical particles appear in realistic gauge quantum field theories in the covariant gauge~\cite{kugo}. The simplest example is the
quantization of the electromagnetic field, where ``time-like'' and ``longitudinal'' photons
do not actually exist, but appear in the intermediate steps of the transition from observable quantities (the vectors $\mathbf E$ and $\mathbf H$) to the
non-observable four-potential $A$, which is made to ensure that the theory is relativistically symmetric and covariant. In order to retain the self-conjugate property for $A$ one has to introduce an indefinite metric in the space of the state amplitudes~(the formal approach of Bleuler~\cite{bleu} and Gupta~\cite{gupt}). In quantum chromodynamics the Faddeev-Popov ghosts are regarded as non-physical particles as well.

In the present paper we consider Haag's theorem in connection with the theories on indefinite metric spaces. Our attention is focused on the class of Krein spaces. The representations of the canonical commutation relations~(the CCR) in Krein space have recently been studied~\cite{jmath,nonph}. In Ref.~\cite{jmath} it is shown that, besides the Fock representation, two other types appear in Krein space, one with negative (the anti-Fock case), the other with two-sided discrete spectrum of the number operator $N=a^+a$. An analogue of the Weyl representation of the CCR for the anti-Fock case in Krein space has been derived in Ref.~\cite{nonph}.

Let us recall the formulation of the generalized
Haag theorem in the standard axiomatic theory~\cite{SW,BLT}. We consider only the case of a neutral scalar field.  
To prove Haag's theorem one must require that the fields smeared in the three space coordinates only,
\begin{equation}
\varphi\,(f,t)=\int\limits_{\mathbb
R^3}\varphi\,(t,x)f(x)\,d^3x,\qquad f\in\mathcal L(\mathbb
R^3),
\end{equation}
are well-defined operators in $\hil$.

\begin{theorem}
Let $\ph_1(f,t)$ and $\ph_2(f,t)$ be two
irreducible sets of field operators at a fixed time $t$ defined
in the Hilbert spaces $\hil_1$ and $\hil_2$, respectively. Let both theories also be invariant with
respect to the proper Poincare group
\begin{gather}
U_j(a,\Lambda)\varphi_j(x)U_j^{-1}(a,\Lambda)=\varphi_j(\Lambda x+a)\label{h1}\notag\\
U_j(a,\Lambda)\Psi_{0j}=\Psi_{0j},\qquad j=1,2\label{h2},
\end{gather}
where $\Psi_{0j}$ is a unique Poincare-invariant (normalized) state in $\hil_j$.
Assume in addition that a unitary transformation $V$
exists that links the fields at an arbitrary time $t$:
\begin{gather}
\varphi_2(f,t)=V\varphi_1(f,t)V^{-1}\label{h3}\notag\\
c\Psi_{02}=V\Psi_{01}\label{h4},
\end{gather}
where c is a complex number with a modulus equal to 1.

If $\ph_1(x)$ is an asymptotic field of mass $m$, then $\ph_2(x)$
is also an asymptotic field of the same mass.
\end{theorem}

The theorem points out the inability of conventional field theories
to describe scattering situations, using the interaction picture, in which the
$S$-matrix is different from the identity.

Most of the proofs of Haag's theorem  are given in the Wightman framework and based on the analytic properties of Wightman functions, which themselves reflect the locality and spectrum conditions. In contrast to this, in Refs.~\cite{Str,Emch} the theorem is proved under more general assumptions, without assuming locality or relativistic invariance. In particular, the proof in Ref.~\cite{Emch} relies on the properties of the general Weyl representation and some other algebraic and group-theoretical facts. In the present paper we are going to apply these arguments and extend Haag's theorem to the theories with non-physical particles, which can be formulated in an indefinite inner product space.

\section{The Weyl Representation of the Canonical Commutation Relations}\label{Weylrep}
\mbox{}\vspace{-\baselineskip}

Let us present a formulation of the Weyl representation in the case of the scalar Bose field according to Ref.~\cite{Emch}.
\begin{definition}
We say that there is a Weyl representation of the canonical commutation relations for the test function space $\mathfrak E$ if
\begin{enumerate}
 \item\label{w1} $\tsp$ is a real pre-Hilbert space;
 \item\label{w2} to each $f$ in $\tsp$ corresponds a pair of unitary operators $U(f)$ and $V(f)$ acting on a Hilbert space $\hil$, common to all $f$ in $\tsp$;
 \item\label{w3} these operators satisfy the following conditions $\forall f_1,\,f_2\in\tsp$:
 \begin{gather}
  U(f_1)U(f_2)=U(f_1+f_2),\notag\\
  V(f_1)V(f_2)=V(f_1+f_2),\notag\\
  U(f_1)V(f_2)=V(f_2)U(f_1)e^{i(f_1,f_2)}\label{gweyl},
 \end{gather}
 \item\label{w4} $U(\lambda f)$ and $V(\lambda f)$ are weakly continuous in $\lambda\in\mathbb R$ for all $f\in\tsp$.
\end{enumerate}
\end{definition}

The connection with a field theory: for each $f$ in $\tsp$ one defines the field
operator $Q(f)$ and its canonical conjugate $P(f)$ as the self-adjoint generators
of $V(\lambda f)$ and $U(\lambda f)$, respectively:
\begin{gather}
V(\lambda f)=\textup{exp}\{-\lambda Q(f)\},\notag\\
U(\lambda f)=\textup{exp}\{-\lambda P(f)\}.
\end{gather}

In Ref.~\cite{Emch} it is shown that  for any finite sequence $\mathcal F =f_1,f_2,\ldots,f_n$ of elements in $\tsp$ there exists in $\hil$ a dense linear manifold $\mathcal D_{\mathcal F}$
stable with respect to $Q(f_j)$ and $P(f_j)$ for all $f_j$ in $\mathcal F$. On $\mathcal D_{\mathcal F}$ the above
composition laws imply for all $f_i,\,f_j$ in $\mathcal F$
\begin{gather}
[Q(f_i),Q(f_j)]=[P(f_i),P(f_j)]=0,\notag\\
[Q(f_i),P(f_j)]=i(f_i,f_j)\ident.\label{gencan}
\end{gather}
The annihilation and creation operators  given by
\begin{gather}
a(f_i)=\frac{\{Q(f_i)+iP(f_i)\}}{\sqrt 2},\notag\\
a^*(f_i)=\frac{\{Q(f_i)-iP(f_i)\}}{\sqrt 2}\label{aa*},
\end{gather}
satisfy the relations
\begin{gather}
[a(f_i),a(f_j)]=[a^*(f_i),a^*(f_j)]=0,\notag\\
[a(f_i),a^*(f_j)]=(f_i,f_j)\ident.
\end{gather}

Now we need to connect the above formalism with the specific case of a free scalar field. The conjugate fields $\phx$ and $\pix=\der\phx/\der t$ satisfy the canonical commutation relations
\begin{gather}
[\phx,\phy]=[\pix,\piy]=0,\notag\\
[\phx,\piy]=i\delta(x-y)\ident.\label{comrel}
\end{gather}
Let us make a transition to a discrete basis~(instead of the continuous $x$ and $y$). For
this purpose we introduce a complete orthonormal set of
functions $\herm$ in three-dimensional space~(e.g., we can choose the Hermite
functions $N_{n}e^{-x^2/2}H_{n_1,n_2,n_3}(x)$) and define the ``coordinates''
and ``momenta'' of the field by
\begin{gather}
Q_n(t)=\int\phx\herm\,d^3x,\notag\\
P_n(t)=\int\pix\herm\,d^3x\label{countbas},
\end{gather}
where $n$ is the compound discrete index $n=(n_1,n_2,n_3)$. In terms of $Q_n$ and $P_n$ the commutation relations \refn{comrel}  take the form
 \begin{gather}
[Q_k(t),Q_l(t)]=[P_k(t),P_l(t)]=0\notag\\
[Q_k(t),P_l(t)]=i\delta_{kl}\ident\label{CCR},
\end{gather}
which is a special case of~\refn{gencan}. The corresponding annihilation and creation operators satisfy
\begin{gather}
[a_k,a_l]=[a^*_k,a^*_l]=0,\notag\\
[a_k,a^*_l]=\delta_{kl}\label{aa}\ident.
\end{gather}

If we define the Weyl operators by
\begin{equation}
U_k(\al)=\exp\{i\al P_k\},\quad V_l(\be)=\exp\{i\be Q_l\},\quad\al,\be\in\re,
\end{equation}
then all the properties~(\ref{w1}-\ref{w4}) will be satisfied for $U_k(\al)$ and $V_l(\be)$. In particular, eq.~\refn{gweyl} will take the form
\begin{equation}
U_k(\al)V_l(\be)=V_l(\be)U_k(\al)e^{i\al\be\delta_{kl}}\label{Weyl}.
\end{equation}
According to Stone~\cite{Stone}, operators $U_k(\al)$ and $V_l(\be)$ are bounded, and hence are defined on the whole Hilbert
space.

The above formulation was given for the operators acting on a Hilbert space. Now we proceed to the consideration of regular representations in a non-degenerate indefinite inner product space.

\section{Krein Space}
\subsection{General properties}\label{krgen}
\mbox{}\vspace{-\baselineskip}

Let us briefly recall some definitions and facts relating to the theory of indefinite inner product spaces. A detailed review is given in Refs.~\cite{Bognar, Aziz, Krein}.
\begin{definition}
Let $\ins$ be an (indefinite) inner product space. The subspace
\begin{equation}
\ins^{\perp}=\{x\in\ins\colon x\perp\ins\}
\end{equation}
is called the isotropic part, and its elements the isotropic vectors of $\ins$.
\end{definition}
Isotropic vectors are orthogonal to all vectors of the space; in particular, they are orthogonal to themselves.
\begin{definition}
We say that the space $\ins$ is non-degenerate if the isotropic part $\ins^{\perp}$ is equal to $0$.
\end{definition}
\begin{definition}
If an inner product space $\kr$ can be represented as the orthogonal direct sum of the form
\begin{equation}\label{dec}
\kr=\kr^+\oplus\kr^-,
\end{equation}
where $\kr^+$ and $\kr^-$ are complete subspaces with positive definite and negative definite metrics, respectively, then we shall say that $\kr$ is a Krein space.
\end{definition}

It is evident that a Krein space is a non-degenerate  inner product space.

According to the definition, the class of Krein spaces includes Hilbert spaces ($\kr^-=0$) as well as anti-spaces of Hilbert spaces ($\kr^+=0$).

Since every vector in a Krein space admits decomposition
\begin{gather}
x=x_++x_-,\,x_{\pm}\in\kr^{\pm},\notag\\
y=y_++y_-,\,y_{\pm}\in\kr^{\pm},
\end{gather}
the inner product of $x$ and $y$ can be written as follows
\begin{equation}
(x,y)=(x_+,x_+)+(y_-,y_-).
\end{equation}
\begin{definition}
We say that $J$ is the fundamental symmetry operator belonging to the fundamental decomposition~\refn{dec} if
\begin{equation}\label{J}
J(x_++x_-)=x_+-x_-.
\end{equation}
\end{definition}
The following properties can easily be verified:
\begin{enumerate}
\item $J$ is completely invertible and $J^{-1}=J$.
\item $J$ is  self-adjoint.
\end{enumerate}

Since $J$ is   self-adjoint, the formula
\begin{equation}
(x,y)_J=(x,Jy)=(x_+,y_+)-(x_-,y_-)
\end{equation}
defines a positive scalar product called the $J$-inner product on a Krein space. Indeed,
$$
(x,x)_J=(x_+,x_+)-(x_-,x_-)\geqslant0.
$$

A norm on $\kr$ can be defined by
\begin{equation}
\|x\|_J=+\sqrt{(x,x)_J},\quad x\in\kr
\end{equation}
and is called the $J$-norm.

It is easy to obtain the relation between the operators $A^*$ and $A^+$, where the symbols ``\,*\,'' and ``\,+\,'' denote adjoint operators relating to the $J$-inner product and the indefinite inner product respectively. Indeed, since
\begin{equation}
(Ax,y)=(Ax,Jy)_J=(x,A^*Jy)_J=(x,JA^*Jy),
\end{equation}
it follows that
\begin{equation}\label{krel}
A^+=JA^*J.
\end{equation}

\subsection{Regular representations in Krein space}
\mbox{}\vspace{-\baselineskip}

Let us return for a moment to the Hilbert space and consider the number operator $N_k=a_k^+a_k$ giving the number of particles in a specific state denoted by the compound index $k$. In section~\ref{Weylrep} quantum field theory was formulated as the quantum mechanics of a system with an infinite number of degrees of freedom, and for a scalar field the index $k$ can be defined via eq.~\refn{countbas}.
Suppose that  $N_k$ has an eigenvector $\Psi_{\alpha_k}$  and
\begin{equation}\label{eigen}
N_k\Psi_{\alpha_k}=\alpha_k\Psi_{\alpha_k}.
\end{equation}
It is known that $\alpha_k\geqslant0$ if $\Psi_{\alpha_k}$ belongs to a Fock space, and,
since $a_k\Psi_{\alpha_k}\sim\Psi_{\alpha_k-1}$, the requirement for $\alpha_k$ to be non-negative imposes  the condition
\begin{equation}
a_k\Psi_0=0\quad\forall k,
\end{equation}
where $\Psi_0$ is a vacuum state.

Condition~\refn{eigen} is analogous to the one that was used as the definition of regularity~\cite{Vernreg} of the CCR representations in Hilbert space in the case of a finite number of degrees of freedom. 

By definition condition~\refn{eigen} introduces regular representations in spaces with an indefinite metric as well. The following theorem classifies the representations of this type in a non-degenerate inner product space~\cite{jmath}:
\begin{theorem}
The regular  representations of the CCR algebra in a non-degenerate inner product space fall into the following classes:
\begin{itemize}
\item Fock case: it is characterized by the existence of a vector $\Psi_0$~(Fock vector) satisfying
$$
a\Psi_0=0.
$$
The number operator $N$ has a complete set of eigenvectors $\Psi_n,\,n=0,1,2,\ldots$, with  non-negative integer eigenvalues;
\item Anti-Fock case: it is characterized by the existence of a vector $\Psi_{-1}$~(anti-Fock vector) satisfying
$$
a^+\Psi_{-1}=0.
$$
The number operator $N$ has a complete set of eigenvectors $\Psi_{n},\,n=-1,-2,\ldots$, with negative integer eigenvalues;
\item $\Lambda$-case: the number operator $N$ has a complete set of eigenvectors such that
$$
\mathrm{Sp}N=\lambda+\mathbb Z,\quad-1<\lambda<0.
$$
\end{itemize}
\end{theorem}
As in subsection~\ref{krgen} we use  the ``+''-notation for the conjugation with respect to the indefinite inner product.

It is important to note that the existence of a regular representation of the CCR in a non-degenerate inner product space implies that the space under consideration is the Krein one. Indeed, consider the set of orthogonal eigenvectors of the number operator. It will be shown below that the eigenvectors can be positive and negative. One can construct a span of all positive~(negative) eigenvectors  and obtain the positive~(negative) definite subspace $\kr^+$~($\kr^-$). Due to the completeness of the set of eigenvectors the whole space can be represented as the direct orthogonal sum of $\kr^+$ and $\kr^-$.

The Fock representations in Krein space are equivalent to those in Hilbert space. The Weyl representation for the CCR algebra was proved to exist in this case~\cite{Vernreg,Foi}. Now we are interested in the anti-Fock case that in our notation implies
\begin{equation}\label{kgr}
a^+_k\Psi_{-1}=0\quad\forall k.
\end{equation}
It is easy to see that
\begin{equation}\label{norm}
(\Psi^{(k)}_{-n},\Psi^{(k)}_{-n})=(-1)^{n-1}\,\,,
\end{equation}
where the set of normalized~(with respect to a single particle space) eigenvectors $\{\psikr_{-n}\}$ of the operator $a^+_ka_k$ is defined by
$$
\psikr_{-n}=\frac{a^{n-1}_k}{\sqrt{(n-1)!}}\,\psikr_{-1},\quad n\in\mathbb N.
$$
Indeed,
\begin{gather}
(\psikr_{-n},\psikr_{-n})=\frac1{n-1}(a_k\psikr_{-n+1},a_k\psikr_{-n+1})=\notag\\
=\frac1{n-1}(\psikr_{-n+1},a^+_ka_k\psikr_{-n+1})=(-1)(\psikr_{-n+1},\psikr_{-n+1})=\notag\\
=\ldots=(-1)^{n-1}(\psikr_{-1},\psikr_{-1}),
\end{gather}
where we can always take $(\psikr_{-1},\psikr_{-1})=1$.

Let us prove that
\begin{equation}\label{antic}
\{a_k,J\}=\{a^+_k,J\}=0\quad\forall k,
\end{equation}
where $\{x,y\}=xy+yx$, is satisfied on a dense domain in Krein space.
Let $n=2m$. Since $a_k\psikr_{-2m}=\psikr_{-2m-1}$, in view of~\refn{J} and~\refn{norm} it is clear that
$$
Ja_k\psikr_{-2m}=\sqrt{2m}J\psikr_{-2m-1}=\sqrt{2m}\psikr_{-2m-1}.
$$
On the other hand,
$$
a_kJ\psikr_{-2m}=-a_k\psikr_{-2m}=-\sqrt{2m}\psikr_{-2m-1}.
$$
The case $n=2m+1$ can be considered similarly. Thus we have shown that
$$
\{a_k,J\}\psikr_{-n}=0,\quad n\in\mathbb N\quad\mbox{and}\quad\forall k.
$$
Next, for every vector $\Psi^{(k)}$ such that
$$
\Psi^{(k)}=\sum_{p=-m}^{-l}c_p\,\psikr_p,\quad m,l\in\mathbb N,\quad m>l\geqslant1,
$$
it follows that
\begin{equation}\label{den}
\{a_k,J\}\Psi^{(k)}=0.
\end{equation}
Thus $\{a_k,J\}=0$ is satisfied on a dense domain in a single particle Krein space. Then, constructing tensor products of $\psikr$ at different $k$, i. e. constructing all possible linear combinations of vectors $a^{n_1}_{k_1}\ldots a^{n_j}_{k_j}\Psi_{-1}$, we extend the latter relation to a dense domain in the entire anti-Fock space. Since $J^+=J$, the relation $\{a^+_k,J\}=0$ holds for all $k$ on the domain as well.

\section{Weyl Representation Analogue. Extension of Haag's Theorem}
\mbox{}\vspace{-\baselineskip}

In this section we show how an analogue of the Weyl representation can be derived for the case of an anti-Fock realization on a Krein space.

Let $a_k$ and $a^+_k$ be the anti-Fock representation operators on a Krein space $\kr$. Let us introduce the operators $b_k=a^+_k$ and $b^+_k=a_k$  and obtain
\begin{gather}
[b_k,b^+_l]=-\delta_{kl}\ident,\quad \tilde{N_k}=-N_k-1,\notag\\
\mathrm{Sp}\tilde{N_k}=\mathbb N,\quad\psikr_{-n}=\tps_{n-1}^{(k)},
\end{gather}
where ``\,\~ \,'' denotes the corresponding notions defined in terms of the new operators.

Relations~\refn{kgr} and~\refn{antic} take the form
\begin{gather}
b_k\tps_0=0\quad\forall k,\notag\\
\{b_k,J\}=\{b^+_k,J\}=0\quad\forall k,
\end{gather}
hence, by~\refn{krel},
\begin{equation}
b^*_k=Jb^+_kJ=-b^+_k.
\end{equation}
Note that the operators $b_k$ and $b^*_k$ satisfy the standard commutation relation of the form~\refn{aa}
\begin{gather}
[b_k,b_l]=[b^*_k,b^*_l]=0,\notag\\
[b_k,b^*_l]=\delta_{kl}\ident.
\end{gather}

In view of~\refn{aa*} the operators $\tp_k$ and $\tq_k$ satisfying the canonical commutation relations~\refn{CCR} can be expressed by
\begin{equation}\label{f}
\tp_k=\frac{(b_k-b^*_k)}{\sqrt2i},\quad\tq_k=\frac{(b_k+b^*_k)}{\sqrt2}.
\end{equation}
It is easy to express them in terms of the original anti-Fock operators as well. Indeed, making use of the relation $b^*_k=-b^+_k=-a_k,\,b_k=a^+_k$, we obtain that
\begin{equation}
\tp_k=\frac{(a^+_k+a)}{\sqrt2i},\quad\tq_k=\frac{(a^+_k-a_k)}{\sqrt2}.
\end{equation}

We see that $\tp_k$ and $\tq_k$ are self-adjoint operators defined on some domain in an effective Hilbert space. Let us introduce the Weyl operators such that
\begin{equation}
U_k(\al)=\exp\{i\al \tp_k\},\quad V_l(\be)=\exp\{i\be \tq_l\},\quad\al,\be\in\re,
\end{equation}

We can repeat the arguments used in Ref.~\cite{Put} in the proof of eq.~\refn{Weyl}. Consequently, $U_k(\al)$ and $V_l(\be)$ satisfy the requirements~(\ref{w1}-\ref{w4}) of the definition in section~\ref{Weylrep}, and we obtain an analogue of the Weyl form of the CCR representation for the anti-Fock case.

Finally, we can apply the corresponding results in Ref.~\cite{Emch} obtained for the Weyl representation  and state that Haag's theorem holds for the case of an anti-Fock representation on a Krein space.

Thus Haag's theorem has been extended to the  theories with non-physical particles.

\end{document}